\documentclass{aa}
\usepackage[varg]{txfonts}
\usepackage[utf8]{inputenc}

\usepackage{graphicx}
\usepackage{subfigure}

\usepackage{natbib}
\bibpunct{(}{)}{;}{a}{}{,} 

\usepackage[uncertainty-mode=separate]{siunitx}
\DeclareSIUnit \pc {pc}
\DeclareSIUnit \erg {erg}
\DeclareSIUnit \msun {{M$_\sun$}}

\newcommand*\diff{\mathop{}\!\mathrm{d}}

\begin{document}
\title{Study of the effect of turbulent interstellar mediums on young supernova remnant morphology}

\author{Gabriel~Rigon\inst{\ref{inst1}}
\and Tsuyoshi~Inoue\inst{\ref{inst1}}\inst{\ref{inst2}}}

\institute{Graduate School of Science, Nagoya University, Furo-cho, Chikusa-ku, Nagoya 464-8602, Japan \email{gabriel.rigon@nagoya-u.jp}\label{inst1} 
\and Department of Physics, Konan University, Okamoto 8-9-1, Higashinada-ku, Kobe 658-8501, Japan \label{inst2} 
}

\date{01 January 0000 /
02 January 0000 }

\abstract{Supernova remnants (SNR) are one of the main sources of galactic cosmic rays' acceleration. This acceleration, believed to happen at the blast wave front, leads to an energy loss at the shock front. This results in the apparent proximity between the blast wave and the contact discontinuity.}
{In this article, we study the effect that turbulent-like density perturbations of the interstellar medium (ISM) have on the evolution of young SNRs. We focus on the impact such fluctuations have on the SNRs' structure and more precisely on the resulting distance between blast wave and contact discontinuity. Since cosmic rays’ acceleration is necessary to explain this distance, this study indirectly put into question the cosmic rays’ acceleration at the blast wave front.}
{We performed a set of purely hydrodynamic three-dimensional simulations without cosmic ray acceleration in a co-expanding frame. We randomly initialised the density variation of the interstellar medium following a Kolmogorov power law. The resulting ratios of radii between blast wave, contact discontinuity and reverse shock are then compared to the astronomical observations.}
{The addition of density perturbation doesn't significantly change the average ratio of radii. However, the simulations show a higher growth of interfacial instabilities in the presence of a turbulent ISM. The resulting deformation of the contact discontinuity could explain the proximity between contact discontinuity and blast wave. They also explain the plateau in the radial distribution of the line of sight velocity associated with the observations of Tycho.}
{The density perturbation of the ISM should not be neglected in the simulation of young SNR as they have an impact comparable to the cosmic rays’ acceleration on the SNR structure.}


\authorrunning{G. Rigon \& T. Inoue}
\titlerunning{Study of the effect of turbulent ISM on young SNR morphology}
\maketitle

\section{Introduction}

Supernova remnants (SNR) are considered a significant actor in the acceleration of galactic cosmic rays (CR) (\cite{blandford_particle_1987}). The observations of nonthermal X-ray synchrotron emission originating from the rims of SNR (\cite{koyama_evidence_1995}) led to the hypothesis of an acceleration of high-energy cosmic ray electrons at a shock front. As CR protons are hard to observe, measuring CR acceleration efficiency is one of the primary interests of this research field. The proximity between blast wave (BW) and contact discontinuity (CD) can be a probe of high-energy CR acceleration efficiency. Indeed, an efficient CR acceleration lowers the post-shock gas pressure thus narrowing the distance between BW and CD. The first analysis of this distance established the BW main as the stage for CR acceleration  (\cite{warren_cosmicray_2005}). An average radius ratio of 1:0.93:0.70 for the respective BW, CD and reverse shock (RS) was measured for the nearly 450 years old type Ia SNR - Tycho (SN 1572). The proximity between BW and CD (0.93) could not be explained by the mono-dimensional (1D) hydrodynamic simulations performed at that time. This led to the current consensus that a significant fraction of the BW energy is lost in CR acceleration.

Since this first work, many studies were performed. Observationally Tycho (\cite{warren_cosmicray_2005,cassamchenai_blast_2007}), SN 1006 (\cite{hamilton_x-ray_1986,cassamchenai_morphological_2008}), Cassiopeia A (\cite{gotthelf_chandra_2001,delaney_kinematics_2004,lee_x-ray_2014}), etc. were studied with similar conclusion. Similarly, more complex simulations were performed to recreate the morphology of these SNRs. The first major process considered was the impact of multi-dimensional simulations on the problem. As it appears in \cite{ferrand_3d_2010,fraschetti_simulation_2010}, the development of interfacial instabilities at the CD, mainly Rayleigh-Taylor instability, leads to greater proximity between CD and BW than predicted with 1D simulations. However, this improvement was not sufficient to explain the actual observations. Other parameters were also studied. For instance, the nature of the supernova (SN) explosion model influences the remnant initial structure and symmetry, which have an important impact on the simulated evolution of the young SNR. In particular, the asymmetry linked to the explosion remains in the remnant for its first few millennia of evolution. Other physical processes, such as the fission of the heavy elements or the presence of a magnetic field (in magneto-hydrodynamic - MHD - simulations), also affect the evolution of the SNR but on a more localised scale (\cite{orlando_fully_2021}). For instance, the fission of the heavy elements leads to a local injection of energy, which results in a quicker expansion of the ejecta in local bubbles (\cite{ferrand_supernova_2019,orlando_fully_2021}). This logically depends on the distribution of these elements, which is asymmetric. The magnetic field tends to stabilise the CD, thus limiting instabilities growth, without significantly changing the relative position of CD and BW.

To this date, the best simulation results, for reproducing the observed radius ratio, were achieved by including the back reaction of CR acceleration on the dynamic of the fluids and shock. In particular, the energy loss at the shock front, due to CR acceleration, should decrease the shock compression ratio. This lower compression ratio results in closer proximity between CD and BW. In order to reproduce this phenomenon in hydrodynamic simulations, it became standard to use an effective adiabatic index for the shocked interstellar medium (ISM) (\cite{ferrand_3d_2010,fraschetti_simulation_2010,orlando_role_2012}). This effective adiabatic index depends on the particle injection rate as well as on the history of the SNR, or more simply the SNR age as shown by \cite{orlando_modeling_2016}'s figure 2. Thus the adiabatic index used in hydrodynamic simulations may evolve from $5/3$ (typical unshocked ISM) to $4/3$ or $3/2$ at the BW front. This artificial correction leads to a higher shock compression and therefore to a thinner shocked ISM shell. While the results obtained through this method are comparatively closer to observations, they still do not reach the right proportion (\cite{ferrand_3d_2010}).

Despite all these considerations on the SNR's initial state and physical mechanisms that may apply, few studies, if any, focus on the effect of the ISM. Usually, spherical symmetry is supposed for the density of the ISM, with either constant or power law, $\rho \propto r^{-2}$, in the case of stellar wind (\cite{ostriker_astrophysical_1988,truelove_evolution_1999,tang_shock_2017}). While such profiles are realistic from an average point of view, as shown in the study of Cassiopeia A by \cite{lee_x-ray_2014}, they do not reflect the reality of the local scales. As the ISM and circumstellar medium are turbulent (\cite{ferriere_plasma_2020}), local density variations are to be expected. This should affect the dynamic of the SNR both in terms of symmetry and BW to CD ratio. Indeed, the propagation of a shock wave inside turbulent medium results in the amplification of the turbulence (\cite{inoue_toward_2012,inoue_origin_2013}) and an energy loss (\cite{chashei_shock_1997,fang_turbulence_2020}). Thus, we can expect the BW to be affected by the turbulence of the ISM in a fashion similar to the feedback of CR acceleration. In addition to these hydrodynamic effects, some MHD effects should also exist, as the compression of a turbulent magnetic field by a shock wave tends to amplify it. We will study this effect in our future works.

In this article, we will study the effect of a turbulent ISM on the evolution of young SNR from a purely hydrodynamic point of view. To limit the effect of external factors on our study, we will neglect most of the physics usually associated with SNR (CR acceleration, asymmetry in the explosion...) and use a simplified profile to initialise our simulations.

This article is organised as follows. In section \ref{method}, we describe our simulation setup. This includes the simulation code we used, the equations that are solved, the initialisation of the turbulence, and the initial profiles of the SNR. In section \ref{results}, we present our analysis method and the results of our simulations, which we compare to SNR observations. Finally, we draw our conclusions in section \ref{conclusion}.

\section{Methods \label{method}}
\subsection{Simulation code}

The starting point of our study is \cite{warren_cosmicray_2005}, which focuses on the nearly 450-year-old type Ia SNR Tycho. In \cite{warren_cosmicray_2005}, a mono-dimensional Lagrangian code was employed. All material was described with a mono-atomic adiabatic equation of state with a constant adiabatic index ($\gamma=5/3$). The simulations were initialised sometime after the SN's explosion ($\sim 2$ weeks), using profiles either from simulations of SN explosions or self-similar SNR expansion models. We should note that each SN explosion model used in \cite{warren_cosmicray_2005} results in a slightly different BW:CD:RS radius ratio, none of which reproduced the observations.

We based our study on the same kind of consideration. We employed the FLASH4 code (version 4.6.2) in a purely hydrodynamic configuration. We used ideal gas equations of state with a single constant adiabatic index equal to $5/3$. Contrary to the above-cited study, we focus on the effect of a turbulent ISM on SNR evolution. Thus, we use three-dimensional simulations (3D). As a result, hydrodynamic instabilities develop at the CD and the BW may be deformed. The initial state of our simulations is also different from Warren's, as our ISM is non-uniform.

Contrary to Warren's Lagrangian simulations, we employed an Eulerian code. This comes with its set of advantages and drawbacks. Eulerian codes are globally simpler in 3D compared to their counterpart. For instance, there is no mesh tangling problem. However, the interfaces between fluids such as the CD are harder to define. In our simulation, we use a passive scalar initialised at the ejecta position to follow this interface. However, this passive scalar tends to diffuse numerically due to the Eulerian nature of the code. This leads to greater uncertainty about the CD position.

To highlight the effect of the turbulence on the SNR evolution, we performed two types of simulations: with a turbulent ISM, or with a uniform ISM, all other parameters being the same. The non-turbulent ISM is used as a reference.

\subsection{Hydrodynamic equation}

One of the characteristics of the SNR evolution is its quick expansion. In 500 years, a young SNR can see its diameter multiplied by a factor of 5000. This brings an important difficulty for Eulerian simulation, where one should balance the simulation's domain size and its actual resolution. As our study deals with the relative position of RS and CD to BW with an expected variation of the order of a few percent (0.86 simulated in \cite{hamilton_x-ray_1986-1} against 0.93$\pm$2 measured by \cite{warren_cosmicray_2005}), we should at least ensure a spatial resolution higher than one 100th of the SNR radius at any time. This can be achieved by a direct approach in 1D, but the computational cost in 3D is too expensive.

The usual approach to keep the total number of cells small would be to employ an adaptive mesh refinement (AMR) scheme (\cite{berger_local_1989,fryxell_flash_2000,orlando_modeling_2016}). However, AMR is not well adapted to simulations of a turbulent domain, as derefinements smooth the turbulence. Furthermore, the numerical cost of an AMR scheme remains important for the late time evolution, as the SNR spans a large part of the domain. This is the reason why we employed the expanding frame approach theoretically developed by \cite{poludnenko_computation_2007}. This approach consists of a change in the referential frame to a co-expanding frame, which is non-inertial. Such an approach has already been employed to study SNR in a uniform ISM with the RAMSES code (\cite{fraschetti_simulation_2010}).

Since the expanding frame is non-inertial, the Euler equation should be modified. In our work, we used a non-rotating non-accelerating expanding frame, with ideal gas equations of state. We set \cite{poludnenko_computation_2007}'s scaling parameters to simplify the equation: $\alpha=\nu=3$ and $\beta=1$ in their notation. As a result, the following equations were numerically solved:
\begin{align}
\partial_\tau \tilde{\rho} + \tilde{\nabla} \cdot (\tilde{\rho} \tilde{\mathbf{u}}) &= 0 \\
\partial_\tau (\tilde{\rho}\tilde{\mathbf{u}}) + \tilde{\nabla} \cdot (\tilde{\rho} \tilde{\mathbf{u}} \otimes \tilde{\mathbf{u}}) + \tilde{\nabla} \tilde{P} &= \mathbf{0}\\
\partial_\tau \tilde{E} + \tilde{\nabla} \cdot(\tilde{\mathbf{u}} (\tilde{E} + \tilde{P})) &= \frac{v_0}{1-v_0 \tau} (2 - 3(\gamma-1)) \tilde{\rho}\tilde{e}\\
\tilde{P} &= (\gamma -1) \tilde{\rho} \tilde{e}
\end{align}
This set of equations corresponds to the Euler equations after applying the following transformation:
\begin{align}
\tilde{\mathbf{r}} &= a^{-1} \mathbf{r} \\
\tau &= \int_0^t \frac{dt'}{a^2(t')} \\
\tilde{\rho} &= a^3 \rho \\
a(t) &= 1 + v_0 t  \quad \text{,} \quad a(\tau) = \frac{1}{1-v_0 \tau} \label{atau}
\end{align}
Here $v_0$ is a constant corresponding to the non-inertial frame expansion rate, $\mathbf{r}$ and $t$ are the space-time coordinate in the inertial frame and $\tilde{\mathbf{r}}$ and $\tau$ are their equivalent in the expanding frame. From this transformation we also have:
\begin{align}
\tilde{P} &= a^5 P\\
\tilde{e} &= a^2 e \\
\tilde{\mathbf{u}} &= a\mathbf{u} - v_0 \mathbf{r} \\
\tilde{E} &= \tilde{\rho}\tilde{e} + \tilde{\rho}\tilde{u}^2/2 
\end{align}
with $\rho$ the density, $P$ the pressure, $\mathbf{u}$ the fluid velocity, and $e$ the internal energy all expressed in the inertial frame. The tilde values are their equivalent in the expanding frame.

Except for the velocity, this transformation does not change the initial profiles of the simulations, since $a(t=\tau=0)=1$. Furthermore, only the energy conservation equation is changed to include a source term. One should however take care as this transformation leads to some other modifications not conspicuous in the above equations. As there is a time-dependent renormalisation of mass, space and time, all physical constants depending on these physical quantities are also time-dependent. For instance in FLASH4, the temperature of the fluids is calculated at each time step using the internal energy: $\tilde{e}=N_a \tilde{k} T/((\gamma-1)A)$, with $N_a$ the Avogadro number, $A$ the average atomic mass, and $\tilde{k}$ the renormalised Boltzmann constant. As $\tilde{k} = a^2 k$, the Boltzmann constant is time-dependent.

The expanding frame also changes the domain boundaries. As the simulation domain expands from the inertial frame point of view, there is an influx of matter from the expanding frame point of view. This influx shapes the ISM in our simulations. Thus, all the information related to the ISM and its turbulence is input into the domain ghost cells at each time step. \cite{poludnenko_computation_2007} highlights that this inflow can result in noise-like features on the velocity field when there is a mismatch between the velocity inputs into the ghost cells and the velocity as extrapolated from the simulation domain. Such features are negligible in our simulations as the expansion rate is constant (no noise was observed in our simulations).

\subsection{Turbulence}

Since we employ an expanding frame to maximise the size of the SNR compared to the simulation domain, the ISM flowing in at the boundary interacts quickly with the BW. The use of a stirring force to generate turbulence is therefore insufficient, as time is required for the turbulence to appear. For this reason, we initialised both our ISM and its inflow with turbulent fluctuations.

In our simulations, only the density field of the ISM is initialised to follow a turbulence spectrum as observed in \cite{armstrong_electron_1995}. The velocity field, which should also present local fluctuations, is taken to be null in the inertial frame. Such a choice is not realistic as there is a correlation between density and fluid velocity. However, the exact details of this correlation are still unknown, so the initialisation of both fields by hand would also bring nonphysical conditions. We considered a constant temperature for the ISM, and we calculate its pressure using the perfect gas equation of state.

We initialised the density in the inertial frame as:
\begin{align}
\rho(\mathbf{r}) &= \rho_0(\mathbf{r}) (1 + \delta \bar{\rho}(\mathbf{r}))
\label{eq_turb1}\\
\text{with} \quad
\delta \bar{\rho}(\mathbf{r}) &= \sum_{\mathbf{k}\neq\mathbf{0}} \sqrt{C k^{-\alpha}} \sin{ \left[ 2\pi \left(\mathbf{k}\cdot\mathbf{r}_L + \Phi (\mathbf{k})\right) \right]} 
\label{eq_turb2}
\end{align}
Here the density, $\rho$, is composed of its average, $\rho_0$, and dimensionless fluctuating part, $\delta \bar{\rho}$. The coefficient of the turbulent power law, $\alpha$, is taken to be equal to $11/3$ to correspond to the 3D Kolmogorov spectrum (\cite{kolmogorov_local_1991}). $\mathbf{r}_L$ is the coordinates renormalised to the field periodicity: $\mathbf{r}_L= \{ x/L_x ; y/L_y ; z/L_z \}$. $\mathbf{k} \in \mathbb{Z}^3$ is a wavevector with a norm $k$. In our simulations, we limit $\mathbf{k}$ to the subset $\llbracket -K; K \rrbracket$ for numerical reasons (here $K=128$). $\Phi$ is a phase, which we randomly initialised.

The implementation of the above formula by a direct calculation is numerically expensive for a 3D domain as the number of operations scale as $\mathcal{O} (K^3N^3)$, with $N$ the number of cells per direction per simulation block. Even, if the initialisation hurdle is overcome similar calculations are needed at each time step to fill the ghost cells. To limit the numerical cost of turbulence initialisation, we based our rationale on the signal processing principle, according to which the quantity of significant data output is equal to the quantity of data input. Accordingly, only a number of points equal to the number of input phases is significant in defining the density variation. Thus, we solve the equation \eqref{eq_turb2} using a fast Fourier transform (FFT) algorithm for a $\mathcal{O} (K^3\log K)$ complexity. As a result of the FFT calculation, we obtain the density variations on a grid $K^3$ cells, which we will thereafter refer to as FFT grid and FFT cells respectively. To obtain the density value on the simulation grid, we linearly interpolate each simulation cell with its neighbours on the FFT grid based on their respective spatial coordinates.

When using an expanding frame, further considerations are required to generate a turbulent inflow of matter. Indeed, the size of the cells varies during the simulations, including the ghost cells. As a result, a simple linear interpolation using the eight neighbouring FFT cells cannot be used. Indeed, as the simulation cells become larger due to the expanding frame, they can contain more than eight FFT cells at late times. This is also due to the fact that the size of the FFT grid cells is constant. The density of a specific simulation cell should thus be the average of the density of each FFT cells it contains, or more precisely a volume-weighted average of the intersection of all FFT cells with this specific simulation cell. We achieved this average through the integration of the FFT cells' values, which we performed before the FFT calculation. Thus, the equation \eqref{eq_turb2} is reformulated as:
\begin{align}
\delta \bar{\rho} (\mathbf{r}) &= \sum H(\mathbf{k}) e^{2\pi i \mathbf{k} \cdot \mathbf{r}_{L}} \label{eq_turb3}\\
H(\mathbf{k}) &= H_0 (\mathbf{k}) H_1 (\mathbf{k}) \\
H_0 (\mathbf{k}) &= -\frac{i}{2} \sqrt{C k^{-\alpha}} \left( e^{2\pi i \Phi(\mathbf{k})} - e^{-2\pi i \Phi(-\mathbf{k})} \right)\\
H_1 (\mathbf{k}) &= \text{sinc}\left(\pi \Delta_x k_x / L_x\right) \text{sinc}\left(\pi \Delta_y k_y / L_y\right) \text{sinc}\left(\pi \Delta_z k_z / L_z\right)
\label{eq_turb4}
\end{align}
where $H(\mathbf{k})$ correspond to the FFT coefficient with $H_0$ directly related to equation \eqref{eq_turb2} and $H_1$ the consequence of the integration. $\text{sinc}$ is the cardinal sine defined as: $\text{sinc}(x)= \sin(x)/x$. And $\Delta_x$ (respectively $\Delta_y$, $\Delta_z$) correspond to width of the cells in the $x$ direction (respectively $y$, $z$) in the inertial frame.

From the algorithmic standpoint, the FFT coefficient and FFT should be recalculated at each time step. The use of an FFT-type algorithm limits the cost of this coherent implementation of turbulence in expanding frame simulations.

The above implementation results in a correct spectrum from the standpoint of the Kolmogorov theory, but also in some nonphysical values. Indeed, $\delta \bar{\rho}$ can reach values lower than $-1$. For this reason, we defined a lower cutoff value, \SI{1e-30}{\g\per\cm\cubed}, to which we artificially set all lower density values.

The strength of the turbulence is set with $C$, however, a singular value of $C$ does not hold a significant meaning without consideration of the distribution of $\delta \bar{\rho}$. This distribution is nearly-Gaussian, has an average of $0$ and its standard deviation, $\sigma$, depends on $C$ and the $256^3$ random phases. In this article, we mainly set $C$ such as $\sigma=0.5$, since this value is realistic in this situation. Indeed, given a density perturbation of the order of the mean density at the driving scale ($L_\text{drive}\sim$\SI{100}{\pc}), and the turbulence scaling as $\ell^{1/3}$, $\ell$ being the spatial scale, we should have density variations of the order of half the average density at around \SI{10}{\pc} (nearly the size of our FFT grid). We also studied some smaller values of $\sigma$ to see the effect of the strength of the turbulence.

\subsection{Simulation parameters}

In our simulations, we work with a Cartesian domain and use the unsplit hydrodynamic solver of FLASH4. We employ an HLLC-type Riemann solver with a second-order spatial reconstruction and a minmod type slope limiter. The simulations are performed on a cubic domain, with an initial length of \SI{2e-3}{\pc}. The majority of the simulations were performed with $300^3$ cells, but some were redone with higher resolution, $520^3$ cells. The simulation domain contains one-eighth of the SNR (an octant). As a result, half of our boundary conditions are reflective (SNR side), while the other half are inflow for the expanding frame (ISM side). The use of reflective boundary conditions allows higher resolution with no additional computational cost. However, artefacts may be produced near these reflective boundaries.

To initialise the SNR, we use a 1D profile supposing a spherical symmetry. The profile is piston-like consisting of radially linear velocity, going from \SI{0}{\cm\per\s} at the centre to \SI{1.2e9}{\cm\per\s} at the CD, and a constant density, $\rho\sim$\SI{9.0e-15}{\g\per\cm\cubed}. The SNR, thus initialised, has a radius of \SI{4.16e15}{\cm}, a mass of 1.37 M$_\sun$ and kinetic energy of \SI{1.16e51}{\erg}. This corresponds to Tycho \SI{e6}{\s} after its explosion according to \cite{badenes_thermal_2003,badenes_thermal_2005}'s simulations, which were used as the initial state for \cite{warren_cosmicray_2005}'s simulations DDTc model. The profile, we used, is simpler than the one resulting from SN explosion simulations, which typically has a non-constant density. As such, our simulations are not meant to recreate the exact state of Tycho, but to understand the effect of the ISM turbulence on the evolution of such SNR. The ISM is initialised with no velocity and an average density of \SI{1e-24}{\g\per\cm\cubed}. The temperature of the ejecta and the ISM is initially set to \SI{10000}{\K}, which is negligible in comparison to the SN kinetic energy. These values for the density and temperature of the ISM correspond to a warm neutral medium.

The expanding rate, $v_0$, was set by taking into account two constraints. First, the whole SNR, BW included, should remain in the simulation domain during the whole simulation. This adds a lower limit to the expanding frame velocity. Second, $v_0$ should not be too large, as the SNR would then contract quickly. This contraction would result in a poorly resolved SNR for the late times. In our simulations, we set $v_0$ to \SI{2.7e-7}{\per\s}. 

We initialise the turbulence with sets of $256^3$ random phases. We use multiple sets to limit the impact a single one would have on our conclusions. We set the periodicity of the turbulence pattern to a single value, $L_x=L_y=L_z$, of the order of the size of the simulation box at the end of the simulation. This limits the smoothing and disappearance of the density fluctuations, that happen when the cell size becomes of the order of $L_x$. We set the periodic length to \SI{8.1}{\pc} for a \SI{8.3}{\pc} simulation domain after 500 years.

\section{Results and analysis \label{results}}
\subsection{Simulations results}
\begin{figure*}
    \centering
    \includegraphics[width=17cm]{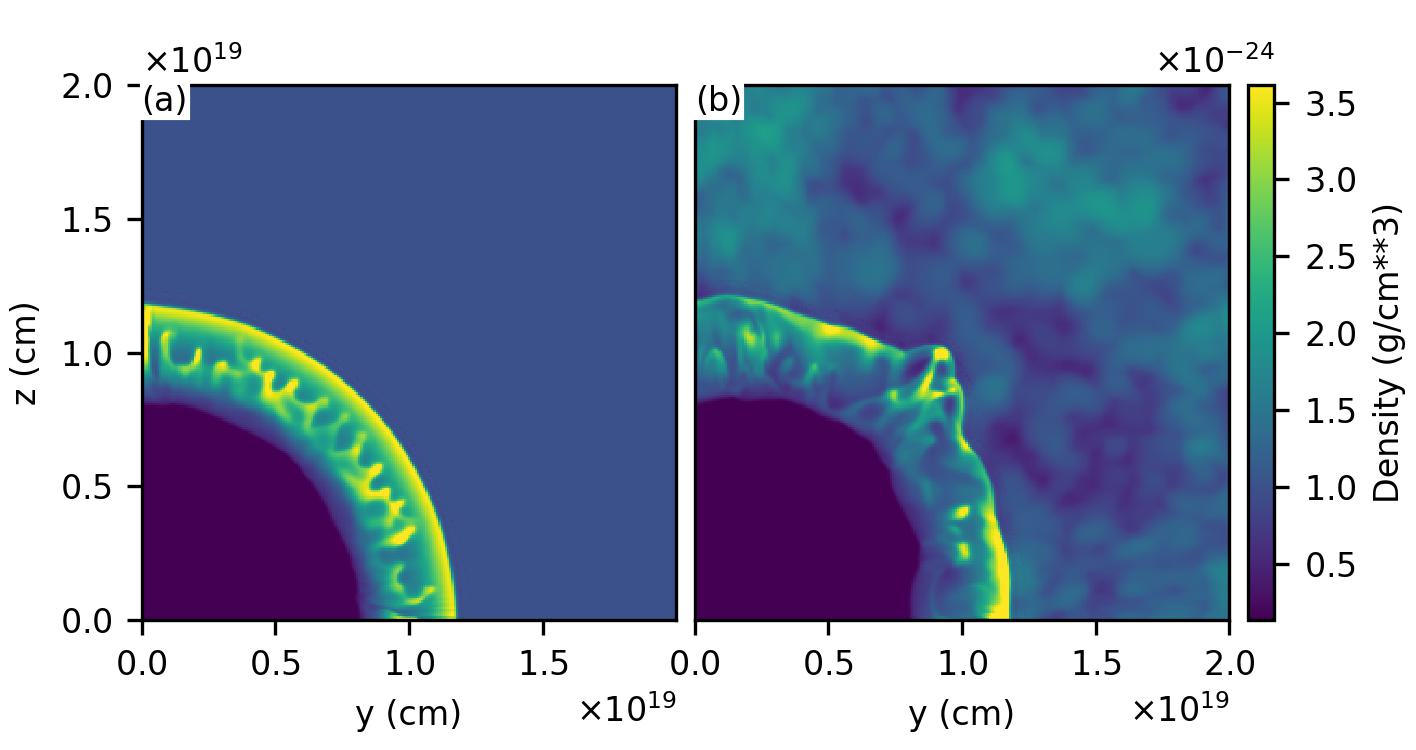}
    \caption{Density slice taken 450 years after SN explosion. The slices are performed on the YZ plane ($x=0$) for both types of simulations: without turbulence (a), and with turbulence (b). The spatial coordinates and density values displayed here are taken in the inertial reference frame. Here, the main effect of the turbulence is the loss of the spherical symmetry of the SNR and a change in the instability wavelength and amplitude at CD.}
    \label{fig:2Dslice}
\end{figure*}

Figure \ref{fig:2Dslice} shows density slices in the YZ plane of the simulations corresponding to a 450-year SNR. In this figure, the displayed values are taken in the inertial frame. As can be seen here, the presence or absence of turbulence has a low impact on the actual density and apparent radius of the remnant. This is not surprising as on average both situations with and without turbulence are the same. However, a major difference exists between both cases. Without turbulence both, BW and RS appear nearly spherical. This is usual in astrophysics as the theoretical answer to a punctual energy perturbation is a spherical shock wave. With turbulence, this spherical symmetry is broken for both BW and RS. This is a direct consequence of the variation of the shock velocity with the density of the medium where it propagates. When the BW encounters an over-density (respectively under-density) of the ISM, its velocity becomes lower (respectively higher). Thus, the shock front gets deformed keeping an imprint of the encountered variations of density. As the BW interacts directly with the density fluctuations of the ISM, its deformations are the most prominent. Whereas the density fluctuations seen by the RS are dampened, as they result from the interaction between ejecta and ISM. The RS deformations are thus more subdued.

Special attention must be paid to the CD as it is not spherical even in absence of turbulence. This deformation of the CD is not a surprise, it is the result of the growth of the Rayleigh-Taylor instability (RTI). This hydrodynamic instability develops itself at the interface between fluids when the gradient of density opposes the inertial pseudo-force in the reference frame of the interface. It appears here due to the deceleration of the CD. This instability cannot be produced in 1D simulations and thus will be a source of divergence with \cite{warren_cosmicray_2005}. In the case without turbulence, the RTI growing at the CD is relatively small and regular. In the turbulent case, the structures created are globally larger and more irregular. A visualisation of the whole simulation shows that deformation of the average position of the interface due to the turbulent variation of density leads to merging between neighbouring spikes or bubbles. Thus, the turbulence affects more than the average position of the CD. We should note that the Richtmyer-Meshkov instability, which should also grow at the CD of young SNR, does not appear in our simulations. This is due to the initialisation of our simulation with the BW outside of the ejecta.

\subsection{Analysis method and interpretation}
\begin{figure}
    \resizebox{\hsize}{!}{\includegraphics{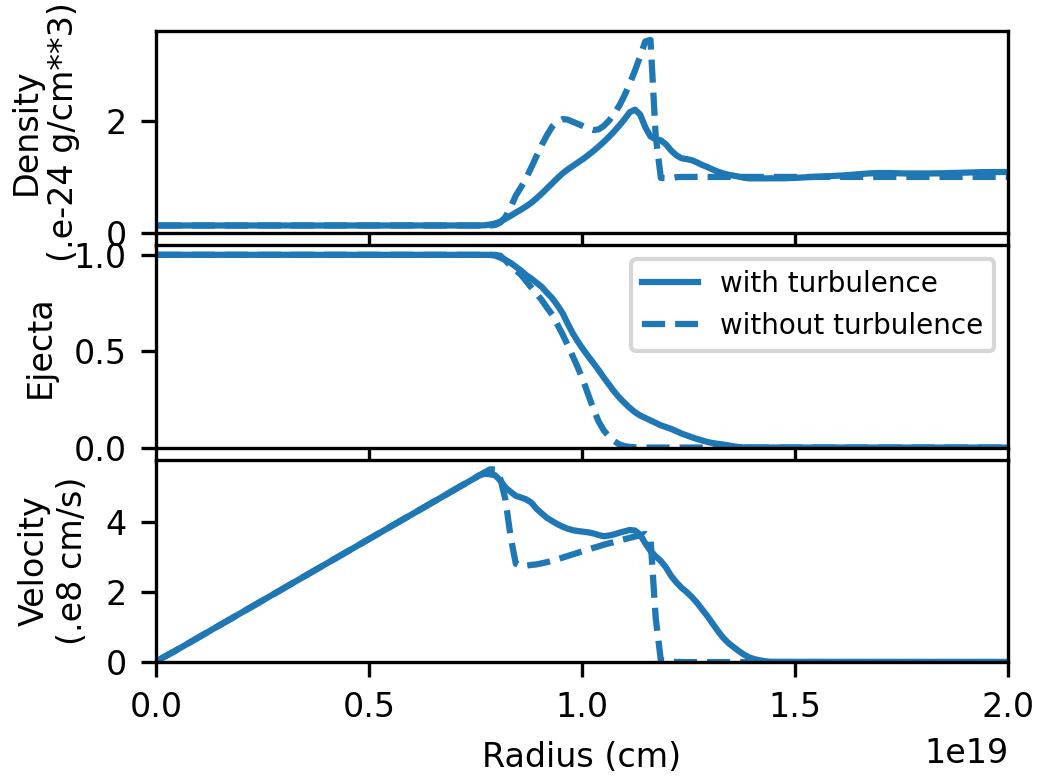}}
    \caption{Radial average profiles 450 years after the SNR explosion. The radial profiles of the density, ejecta and norm of the velocity are displayed in the inertial coordinates both with (plain curve) and without (dashed curve) turbulent ISM. These profiles are averaged over all angles of the octant.}
    \label{fig:RadialAverage}
\end{figure}

Figure \ref{fig:RadialAverage} shows the density, velocity and ``ejecta'' radial profiles associated to figure \ref{fig:2Dslice}, they are obtained by performing an average over all angles. Here, the ``ejecta'' field corresponds to a passive scalar initialised at the position of the ejecta material. This passive scalar is transported by the fluid and is thus subject to both convection and numerical diffusion, but not to physical diffusion. A value of 1 corresponds to \SI{100}{\percent} ejecta and 0 to \SI{100}{\percent} ISM, the value in between showing a mix between both materials. In the absence of turbulence, the 1D profiles in figure \ref{fig:RadialAverage} give us a good idea of the positions of the three interfaces, we are interested in. The discontinuities in velocity and density correspond to the BW and RS. The CD position is usually taken at the local density minimum between both shocks (\cite{fraschetti_simulation_2010}), in our case, this corresponds to an ejecta's value of 0.14. Hereafter we will use 0.1 as the value for the CD position. In presence of turbulence, the loss of spherical symmetry leads to a broadening of all structures and a smooth transition in density at the CD. The shocks might appear broader in this figure, however, this is due to the spatial average and not to a physical broadening of these shocks.

\begin{figure}
    \resizebox{\hsize}{!}{\includegraphics{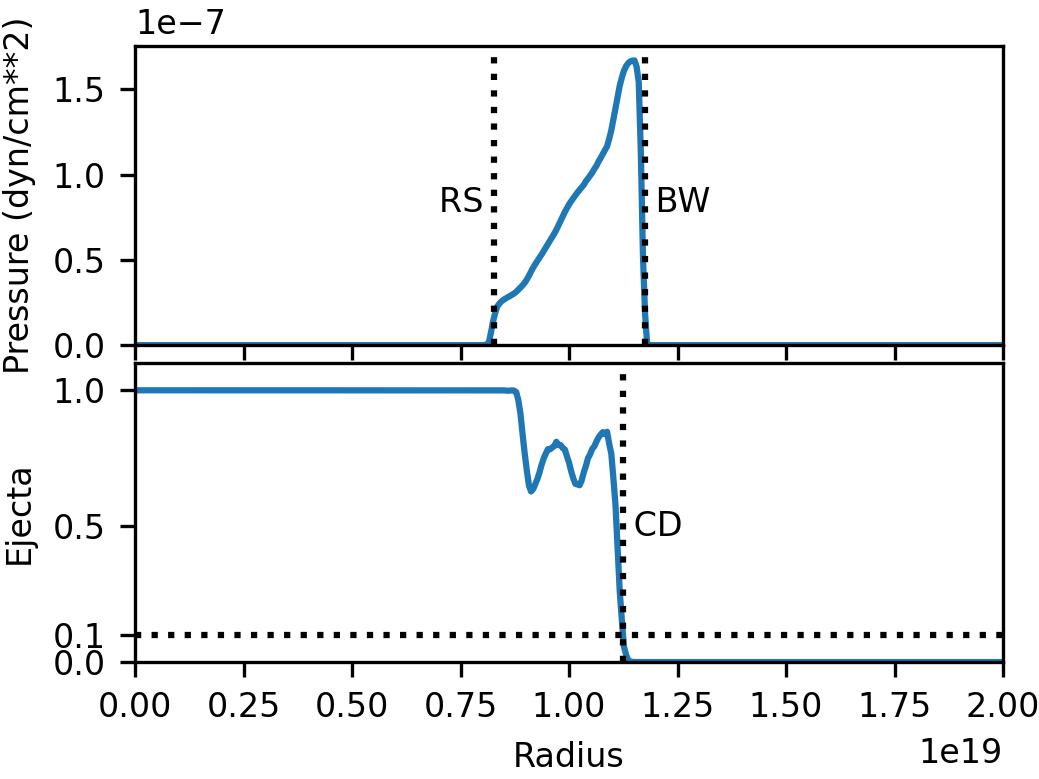}}
    \caption{Radial profiles for an arbitrary solid angle 450 years after SNR explosion. The pressure and density profiles of the simulation with turbulence are displayed. The discontinuities in the pressure profile correspond to the BW and RS. The CD is defined as the point, where the ejecta crosses 0.1.}
    \label{fig:SolidAngle}
\end{figure}
To perform a quantitative study of our simulations and compare our results to Warren's observations, we have to consider the relative position of BW:CD:RS in each spatial direction. For this, we project our Cartesian simulation on a Spherical covering grid and then find the radius of the BW, CD and RS for each solid angle. Figure \ref{fig:SolidAngle} shows a line-out of the pressure and the ``ejecta'' for an arbitrary taken solid angle. The two jumps in pressure correspond to the BW and RS, and a value of 0.1 in the ``ejecta'' field corresponds to the CD. In our analysis, we used the pressure field to find both BW and RS, as the simplicity of this profile makes automating their measure easier. We should add that in some situations, the position of each interface was not unique for some solid angles. This happens when the tip of RTI spikes broadens in the non-linear phase of the instability. We removed these points from the analysis present hereafter as their interpretation is ambiguous. However, these points only have a small effect on the following analysis. For instance, for the 450-year SNR with a turbulent ISM, the ratio of radius between CD and BW varies between \num{0.874(32)} when keeping these points and \num{0.872(30)} when removing them.

\begin{figure*}
    \sidecaption
    \includegraphics[width=12cm]{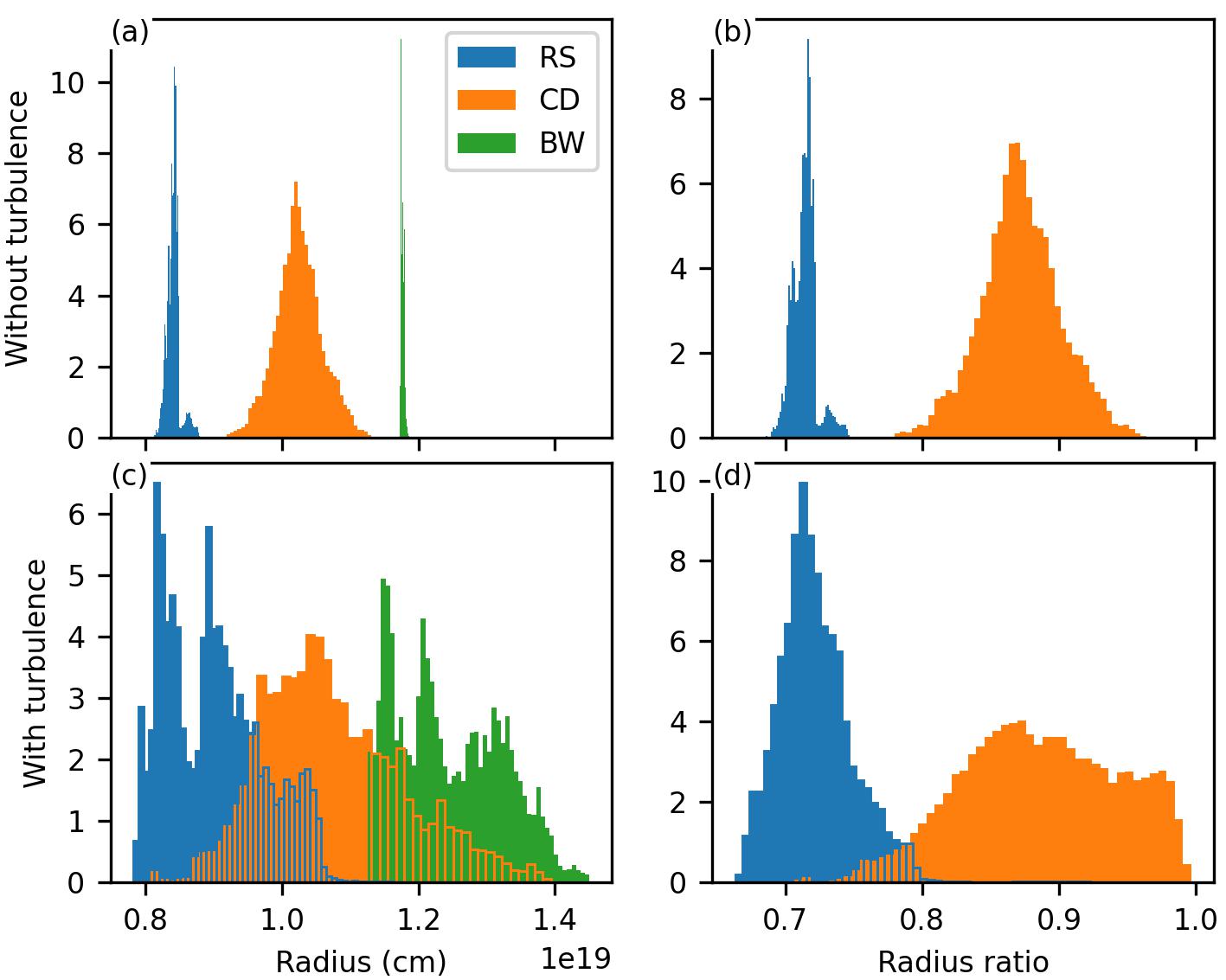}
    \caption{Spatial distribution of the SNR radius. Histogram of the RS (blue), CD (orange), BW (green) radius both without (a) and with (c) turbulence obtains 450 years after SN explosion (simulations displayed on Fig.\ref{fig:2Dslice}). To compensate for the variation spatial variation of BW radius appearing in the turbulent case, the RS and CD radii are normalised using their respective BW radius for each solid angle both without (b) and with (d) turbulence.}
    \label{fig:Distribution}
\end{figure*}
From the analysis method described above, we can associate each time step of the simulations with the distributions of BW, CD, and RS radii. Figure \ref{fig:Distribution} shows the distributions associated with the figure \ref{fig:2Dslice} simulations. As can be seen, without turbulence the BW and RS both have a sharp distribution (small standard deviation \SI{\sim 3e16}{\cm} for the BW) and thus are spherical. The CD have a broad distribution, with a standard deviation of \SI{3.6e17}{\cm} for a mean value of \SI{1.021e19}{\cm}. This is due to the RTI growth. Here the standard deviation of the CD distribution is a good indication of the scale of the mixing zone width, which is a source of uncertainties in astronomical observations. In the turbulent case, all distributions are broader due to the loss of spherical symmetry.

By normalising the interface radius by either the BW or the RS radius, we can partially compensate for the loss of spherical symmetry. The resulting distributions shown in the right panels of figure \ref{fig:Distribution} are still broader in the turbulent case as the BW is more sensible to ISM turbulence than the RS and CD as explained previously.

\subsection{Aspect ratio and RTI}
\begin{figure}
    \resizebox{\hsize}{!}{\includegraphics{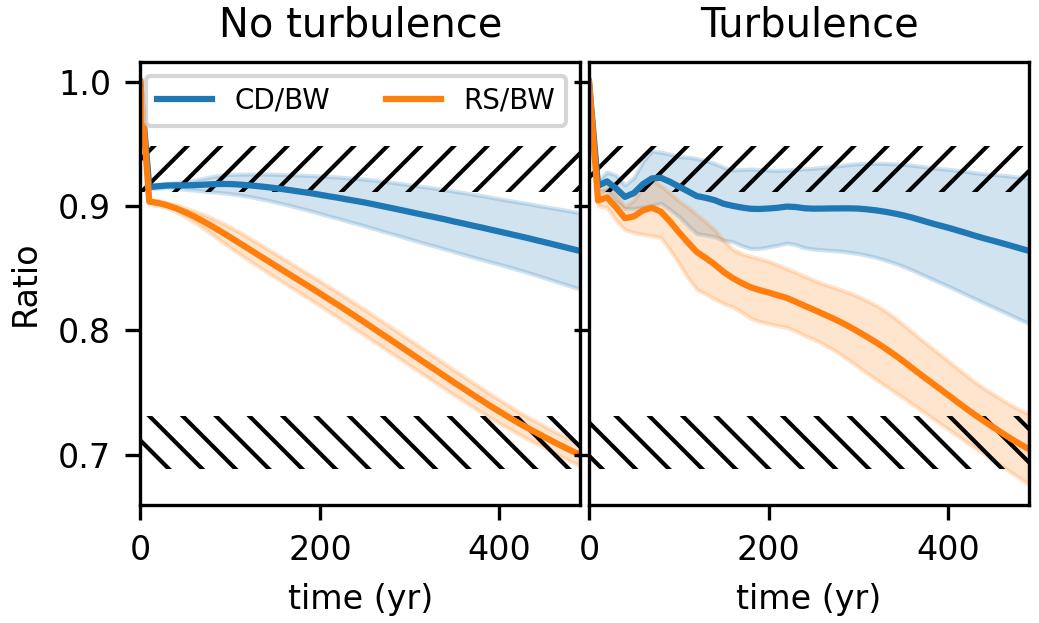}}
    \caption{Time evolution of the various ratios of BW, CD and RS' radii. The temporal evolution of the CD (blue) and RS (orange) radii normalised by the BW radius is displayed for 500 years both for the simulation without (left) and with (right) turbulence. The plain curve corresponds to the average over all solid angles of the distribution of radii ratio, and the coloured zones to their standard deviation. The hatched areas correspond to the radius ratios as observed in \cite{warren_cosmicray_2005}.}
    \label{fig:TEvol}
\end{figure}
Considering the normalised distribution obtained at each time step we can calculate the temporal evolution of the ratios of BW:CD:RS and compare them to Warren's observation. Figure \ref{fig:TEvol} shows the temporal evolution of the average of the distributions of each ratio. The coloured zones correspond to the standard deviation of each distribution, which is due either to RTI growth and/or to partially compensated loss in spherical symmetry. The hatched zones correspond to the ratios observed by Warren. In a similar fashion as in \cite{warren_cosmicray_2005}, the observed BW:CD and BW:RS ratios are not reproduced simultaneously by our non-turbulent simulation. This is also the case for simulations with turbulent ISM when working only with average, however, the conclusion is reversed when taking into account the standard deviation. From this two conclusions can be drawn. First, the ISM turbulence does not significantly change the radii ratios on average. So in purely hydrodynamic simulations, turbulent density variations of the ISM and the resulting post-shock turbulence do not significantly change the effective adiabatic index. However, with turbulence, the RTI growing at the CD becomes large enough to reproduce the observed ratios. This brings into question the true nature of Warren's observations. If their observations have a high enough resolution to see the deformation of the interface due to RTI and if their analysis method correctly compensates for the 3D nature of Tycho, then the effect of the ISM turbulence alone is insufficient. However, if the resolution of the observation is too low, a bias in the measure of the CD radius might appear, shifting the CD's measured position in the direction of the RTI spikes. In this case, the ISM turbulence can explain the observation.

\begin{figure}
    \resizebox{\hsize}{!}{\includegraphics{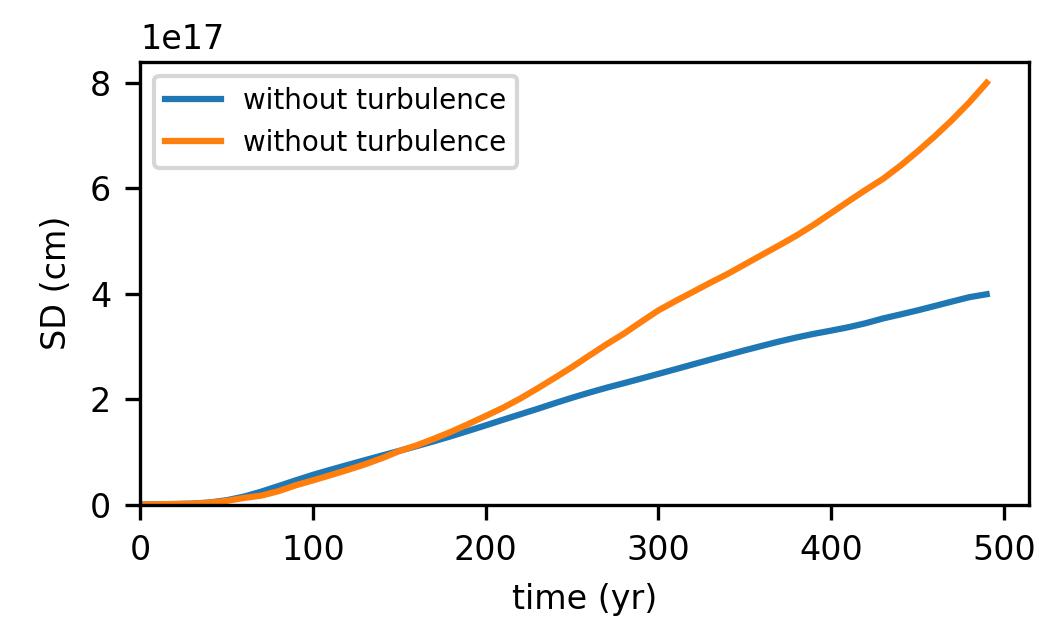}}
    \caption{Time evolution of the standard deviation of the CD after spherical compensation. The values displayed here correspond to the standard deviation of the distribution of CD/RS radii multiplied by the average RS radius. The RTI at the CD and thus the mixing zone has a quicker growth in a turbulent ISM (orange) than in a non-turbulent one (blue).}
    \label{fig:RTI}
\end{figure}
Figure \ref{fig:RTI} shows the temporal evolution of the CD standard deviation after spherical compensation by the RS radius (the CD radius obtained for each solid angle is divided by the equivalent RS radius and multiplied by the average RS radius $\langle\frac{r_\text{CD}}{r_\text{RS}}\rangle\langle r_\text{RS}\rangle$). As can be seen in this figure, the RTI has a quicker growth in presence of turbulence. The temporal evolution of the CD average position is similar in both cases, with a slightly higher deceleration in absence of turbulence. Thus, the difference in RTI growth is not due to differences in acceleration profiles of the interface, but mainly to the deformation of RTI (change in wavelength, merging of structures, etc) and to the change in ambient medium (variation of Atwood number due to the varying ISM density and zone of higher and lower pressure due to the deformation of the shock).

This enhancement in RTI growth may remind us of previously performed 2d studies of the interaction of young SNR with clumpy circumstellar medium (\cite{jun_interaction_1996}). In these studies, the interaction of the BW with spherical clumps of over-densities present in the ISM results in the creation of eddies. Contrarotative eddies form flow channels that boost the kinetic energy of well-placed RTI fingers, and thus their growth. Our simulations show that the average vorticity of the shocked ISM is four times higher in the turbulent case compared to the non-turbulent one. However, rather than being well distributed in the shocked shell, the high vorticity regions are concentrated along the RTI spikes at CD. This is particularly clear for the non-turbulent case where the gap between BW and CD is devoid of vorticity, while high-vorticity zones clutter the RTI. This suggests that rather than being created by the interaction between the BW and the density perturbations, the eddies are due to the growth of the RTI [\cite{bian_revisiting_2020,rigon_exploring_2021}]. As mentioned before, in our case, the perturbations of density influence the RTI mainly through the changes in Atwood number and through the deformation of the BW, which results in high and low-pressure regions (cf. mechanism of the Richtmeyer-Meshkov instability). Contrary to the studies of clumps, in presence of turbulence, the changes in density are progressive (clumps have an interface with the ISM) and the post-shock eddies might not align in 3d as well as in 2d. 

\subsection{Discussion and limits}
The turbulence present in our simulations enhances the growth of RTI and leads to BW:CD:RS ratios, which can be compared to Warren's observations. This conclusion does not depend on the specific simulation shown in this article. We performed additional simulations with different random sets of phases for the turbulence (see appendix for additional curves). In all of them, the RTI growth is enhanced by the turbulence and a temporal range where both CD/BW and RS/BW ratios are reproduced simultaneously can be found. However, the actual temporal range may be shifted (with some as early as 370 years) and the RTI growth, while keeping the same concavity, slightly differs from case to case.

However, we should note that the strength of the turbulence is an important parameter. For instance, for $\sigma=0.05$, so for a ten-time lower density perturbation, the simulation results are close to the case without turbulence. In particular, both the BW and RS are nearly spherical. The standard deviation of the BW radius corresponds, in this case, to \SI{0.9}{\percent} of the average BW radius against \SI{0.2}{\percent} without turbulence. Furthermore, both BW:CD and BW:RS ratios are not reproduced simultaneously, as in the no-turbulence case.

As in any other numerical study of the young SNR, the initial profile of the remnant plays an important role in its dynamic, even when neglecting all initial asymmetry. Thus, one can wonder if our conclusion still held true with more realistic initial profiles. To bring a tentative answer to this question, we performed a pair of simulations, with and without turbulent ISM, starting from the DDTc radial profile used in \cite{warren_cosmicray_2005}. This profile presents in particular a drop in density near the edge of the remnant (CD) and two pressure spikes near its centre. The dynamic of the resulting remnants is different from our piston case, and it requires a higher expansion rate (\SI{9.2e-7}{\per\s}). However, similar conclusions on the RTI growths and radii ratios can be drawn. More precisely, in the simulation, we performed based on a DDTc profile, the BW:CD:RS ratios could not be recreated on average (without taking into account the RTI) but are obtained only in the turbulent case when taking into account the standard deviation. In the specific simulation performed, both ratios were obtained simultaneously between 332 and 423 years.

\subsection{Qualitative comparison to observations}
Our simulations can also be qualitatively compared to SNR observations. While such comparisons are more limited as they are based on subjective factors, due to the lack of metrics, they can still give us some ideas of global trends.

A direct observation of the results of our simulations, such as the one displayed in figure \ref{fig:2Dslice}, already contains numerous information on the remnants and more precisely on their morphology. We already mentioned that the remnant stays nearly spherical with the exception of the CD in the non-turbulent case. To this, we should add that, while RTI grows at the CD, most of it keeps a limited size. As a result, there is a gap between RTI spikes and the BW. On the other hand, in presence of a turbulent ISM, the remnant is deformed as a whole. The RTI is less regular, and we can observe long deformed spikes that reach the BW and even push it. This results in protrusions of the BW, which don't exist in the non-turbulent case. These deformations (loss of spherical symmetry and protrusions) can be found in real systems such as Tycho. This can be seen as another proof of the importance of the ISM's perturbation on the SNR evolution.

We should also mention here the limit of our simulations. It is well known that the BW of an SNR is non-collisional since the plasma's conditions estimated at this interface suggest a mean-free path of the order of the radius of the SNR or even more. Our simulations are based on a hydrodynamic code FLASH4. As such, all shocks present here are collisional. This qualitative difference between code and observation might not be obvious in our displays, but it constitutes a major limitation of our work even if the trends are the same(deformation of the BW, enhancement of the RTI growth).

\subsection{Comparison to observed velocities}

A recent analysis of observations of Tycho results in the measure of radial profiles of the line of sight (LoS) velocity (\cite{kasuga_observational_2021}). This analysis is based on the Doppler shift of line emission of the heavy elements of the SNR. It shows a drop in the LoS velocity, that can be approximated by $\sqrt{(R-r)/(R+r)}$, followed by a plateau near the edge of the SNR. While the first part can be understood and modelled by simple spherical shell-like emission, the plateau is controversial as it cannot be recreated by this simple model. \cite{kasuga_observational_2021} suggested that this plateau feature may be due to the interaction of the SNR with a dense medium.

\begin{figure}
    \resizebox{\hsize}{!}{\includegraphics{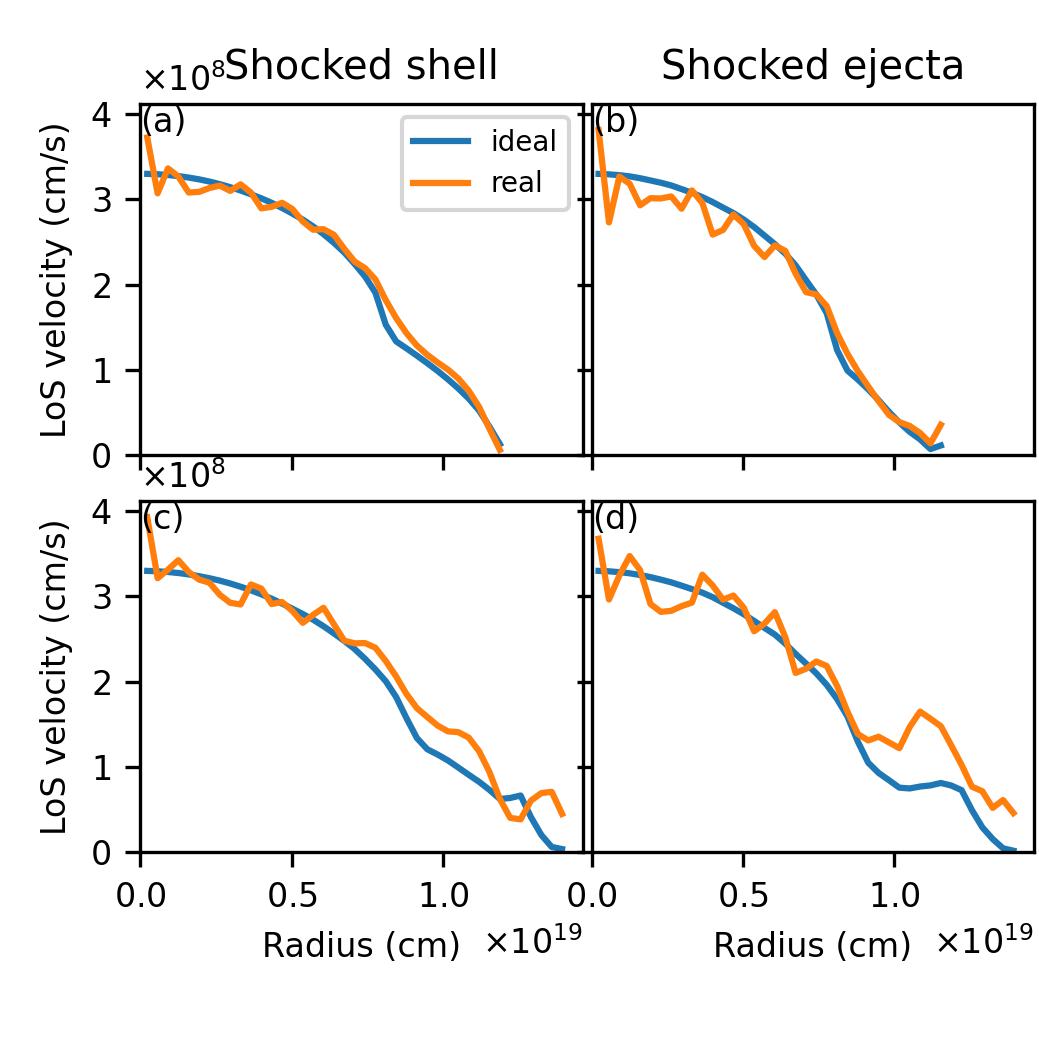}}
    \caption{Line of sight velocity 450 years after SN explosion. The average density-weighted line of sight (LoS) velocity is displayed both for the non-turbulent (a,b) and turbulent (c,d) cases. The calculation of the LoS velocity uses either the shocked shell contained between BW and RS (a,c) or the shocked ejecta contained between CD and RS (b,d). Both idealised values (blue) and simulated values (orange) of the velocity and density are considered for the calculation.}
    \label{fig:LoS}
\end{figure}

By post-processing our simulations, we can study the effect of turbulence on the radial profile of the LoS velocity. Figure \ref{fig:LoS} shows different radial profiles depending on the simulation (with/without turbulence) and on the calculation method of the density averaged LoS velocity. To calculate the LoS we consider here either the shocked shell or the shocked ejecta as the emitting medium. The shock shell is nearly equivalent to the models developed by \cite{kasuga_observational_2021}, where a spherical shell emits the light. However, the shocked ejecta is according to us more relevant. Only it contains the heavy elements producing the line emission, whose Doppler shifts are observed. The results shown in figure \ref{fig:LoS} use both idealised values and real values for the velocity and density. The idealised values suppose a constant radial velocity and a constant density. The real values use the results of the simulation. 

As can be seen in the figure, idealised and real values bear similar results, with noisier profiles for the real values. This means the exact velocity and density distributions are close to their idealised value and that they have a low impact on the observed LoS velocity. On the other hand, the nature of the emitters has an important impact on the profile. For instance, in the case of idealised values with the shocked shell and without turbulence, the LoS velocity profile obtained corresponds to the results of \cite{kasuga_observational_2021} model, but is inconsistent with their observations, as there is no plateau at the edge of the profile. The radial profile is, however, completely changed when considering only the shocked ejecta. In particular, a plateau appears at the SNR edge in the turbulence case. This plateau is an indirect result of the RTI growth, which shapes the ejecta boundary. The regularity of this instability in the non-turbulent case leads to the quasi-linear drop in LoS velocity. In the case of turbulence, the loss of regularity leads to the superimposition of large and small spikes and as a result, a plateau may be created.

From the profiles we obtained, we can conclude that the ISM turbulence can lead to the creation of a plateau and that the LoS velocity observed corresponds to the ejecta. In this case, the beginning of the plateau nearly coincides with the base of the bubble and its end to the tip of the RTI spikes. As such, the observation of this plateau might be a good measure of the RTI growth and resulting mixing zone at the CD.

\section{Conclusions \label{conclusion}}

In conclusion, we performed 3D hydrodynamic simulations of the evolution of a young SNR, initialised on conditions supposed to correspond to Tycho. These simulations employed only a constant adiabatic index, contrary to the usual SNR simulation using an effective adiabatic index to take into account the effect of CR acceleration. The main aim of these simulations is to explore the effect that turbulent-like density variations of the ISM would have on the SNR dynamic.

We showed in this article, that a turbulent ISM only has a small effect on the average position of BW, RS and CD. However, it leads to a loss in the spherical symmetry of the remnant and a higher RTI growth at CD. This faster-growing RTI might explain the aspect ratio of BW:CD:RS measured in \cite{warren_cosmicray_2005}. It also seems responsible for the plateau observed on the radial profiles of the line of sight velocity (\cite{kasuga_observational_2021}). Thus, the measure of this plateau is a good indication of the mixing zone width linked to the RTI growth.

\begin{acknowledgements}
G. Rigon's work was performed as an international research fellow of the Japan Society for the Promotion of Science (postdoctoral fellowship for research in Japan (Standard)).  This work is supported by Grants-in-aid from the Ministry of Education, Culture, Sports, Science, and Technology (MEXT) of Japan (20H01944). Simulations were performed on the XC50 supercomputer of the Center for Computational Astrophysics of the National Astronomical Observatory of Japan, under XC-trial and XC-B type proposals.
\end{acknowledgements}

\bibliographystyle{aa}
\bibliography{references}

\begin{appendix}
\onecolumn
\section{Deriving the equations on turbulence}
As the relation between the equation \eqref{eq_turb2} and the set of equations from \eqref{eq_turb3} to \eqref{eq_turb4} might not be obvious, we wrote here the detail of this transition.

\begin{align*}
\delta \bar{\rho}(\mathbf{r}) &= \sum_{\mathbf{k}\in \llbracket -K; K \rrbracket^3;\mathbf{k}\neq\mathbf{0}} \sqrt{C k^{-\alpha}} \sin{ \left[ 2\pi \left(\mathbf{k}\cdot\mathbf{r}_L + \Phi (\mathbf{k})\right) \right]} \\
&= \sum_{\mathbf{k}\in \llbracket -K; K \rrbracket^3;\mathbf{k}\neq\mathbf{0}} \sqrt{C k^{-\alpha}}
\left(
e^{ 2i\pi \left(\mathbf{k}\cdot\mathbf{r}_L + \Phi (\mathbf{k})\right) }
- e^{  -2i\pi \left(\mathbf{k}\cdot\mathbf{r}_L + \Phi (\mathbf{k})\right) }
\right)/2i\\
&= \sum_{\mathbf{k}\in \llbracket -K; K \rrbracket^3;\mathbf{k}\neq\mathbf{0}} \sqrt{C k^{-\alpha}}
e^{ 2i\pi \mathbf{k}\cdot\mathbf{r}_L}\left(
e^{ 2i\pi \Phi (\mathbf{k}) }
- e^{  -2i\pi \Phi (\mathbf{-k}) }
\right)/2i\\
&= \sum_{\mathbf{k}\in \llbracket -K; K \rrbracket^3;\mathbf{k}\neq\mathbf{0}} H_0(\mathbf{k})e^{ 2i\pi \mathbf{k}\cdot\mathbf{r}_L}
\end{align*}
In this last form, we can recognised a discreet Fourier transform with Fourier coefficients equal to $\sqrt{C k^{-\alpha}}\left(e^{ 2i\pi \Phi (\mathbf{k}) }- e^{  -2i\pi \Phi (\mathbf{-k}) }\right)/2i$. An FFT-type algorithm can be used to solve it.

Now, let's consider the average density, $\langle\rho\rangle$ of a simulation cell with centre $(x_0, y_0, z_0)$ and size $(\Delta_x, \Delta_y, \Delta_z)$.
\begin{align*}
\langle\rho\rangle(x_0, y_0, z_0) &= \int_{x_0-\Delta_x/2}^{x_0+\Delta_x/2}\int_{y_0-\Delta_y/2}^{y_0+\Delta_y/2}\int_{z_0-\Delta_z/2}^{z_0+\Delta_z/2} \frac{\rho(x, y, z)}{\Delta_x \Delta_y \Delta_z} \diff x\diff y \diff z \\
&= \rho_0 \left(1 +\int_{x_0-\Delta_x/2}^{x_0+\Delta_x/2}\int_{y_0-\Delta_y/2}^{y_0+\Delta_y/2}\int_{z_0-\Delta_z/2}^{z_0+\Delta_z/2} \frac{\delta \bar{\rho}(x, y, z)}{\Delta_x \Delta_y \Delta_z} \diff x\diff y \diff z \right)\\
&= \rho_0 \left(1 + \sum_{\mathbf{k}\in \llbracket -K; K \rrbracket^3;\mathbf{k}\neq\mathbf{0}} \frac{H_0(\mathbf{k})}{\Delta_x \Delta_y \Delta_z}
\int_{x_0-\Delta_x/2}^{x_0+\Delta_x/2}\int_{y_0-\Delta_y/2}^{y_0+\Delta_y/2}\int_{z_0-\Delta_z/2}^{z_0+\Delta_z/2}
e^{ 2i\pi (k_x x/L_x +k_y y/L_y + k_z z/L_z )} \diff x\diff y \diff z
\right)\\
&= \rho_0 \left(1 + \sum_{\mathbf{k}\in \llbracket -K; K \rrbracket^3;\mathbf{k}\neq\mathbf{0}} \frac{H_0(\mathbf{k}) L_x L_y L_z}{(2i\pi)^3\Delta_x \Delta_y \Delta_z k_x k_y k_z}
\left(e^{ 2i\pi k_x (x_0 + \Delta_x/2)/L_x}- e^{ 2i\pi k_x (x_0 - \Delta_x/2)/L_x}\right)\right.\\
&\qquad \left. \vphantom{\sum_{\mathbf{k}}}
\left(e^{ 2i\pi k_y (y_0 + \Delta_y/2)/L_y}- e^{ 2i\pi k_y (y_0 - \Delta_y/2)/L_y}\right)\left(e^{ 2i\pi k_z (z_0 + \Delta_z/2)/L_z}- e^{ 2i\pi k_z (z_0 - \Delta_z/2)/L_z}\right)
\right)\\
&= \rho_0 \left(1 + \sum_{\mathbf{k}\in \llbracket -K; K \rrbracket^3;\mathbf{k}\neq\mathbf{0}} \frac{H_0(\mathbf{k}) L_x L_y L_z}{(2i\pi)^3\Delta_x \Delta_y \Delta_z k_x k_y k_z}
e^{ 2i\pi k_x x_0/L_x}(2i \sin{2\pi k_x \Delta_x/2L_x}) \right.\\
&\qquad \left. \vphantom{\sum_{\mathbf{k}}}
e^{ 2i\pi k_y y_0/L_y}(2i \sin{2\pi k_y \Delta_y/2L_y})e^{ 2i\pi k_z z_0/L_z}(2i \sin{2\pi k_z \Delta_z/2L_z})
\right)\\
&= \rho_0 \left(1 + \sum_{\mathbf{k}\in \llbracket -K; K \rrbracket^3;\mathbf{k}\neq\mathbf{0}} H_0(\mathbf{k}) 
e^{ 2i\pi \mathbf{k}\cdot\mathbf{r}_L}\text{sinc}(\pi k_x \Delta_x/L_x)\text{sinc}(\pi k_y \Delta_y/L_y)\text{sinc}(\pi k_z \Delta_z/L_z)
\right)\\
&= \rho_0 \left(1 + \sum_{\mathbf{k}\in \llbracket -K; K \rrbracket^3;\mathbf{k}\neq\mathbf{0}} H_0(\mathbf{k}) H_1(\mathbf{k}) e^{ 2i\pi \mathbf{k}\cdot\mathbf{r}_L}
\right)
\end{align*}
We retrieve here the formula used in this article to generate the turbulence. The $H_1$ part depends on the size of the simulation cell and thus evolves with time.

\section{Additional plots}
\begin{figure}
\centering
\subfigure[Evolution of the ratios of radii\label{fig:TEvolb}]{
\includegraphics[]{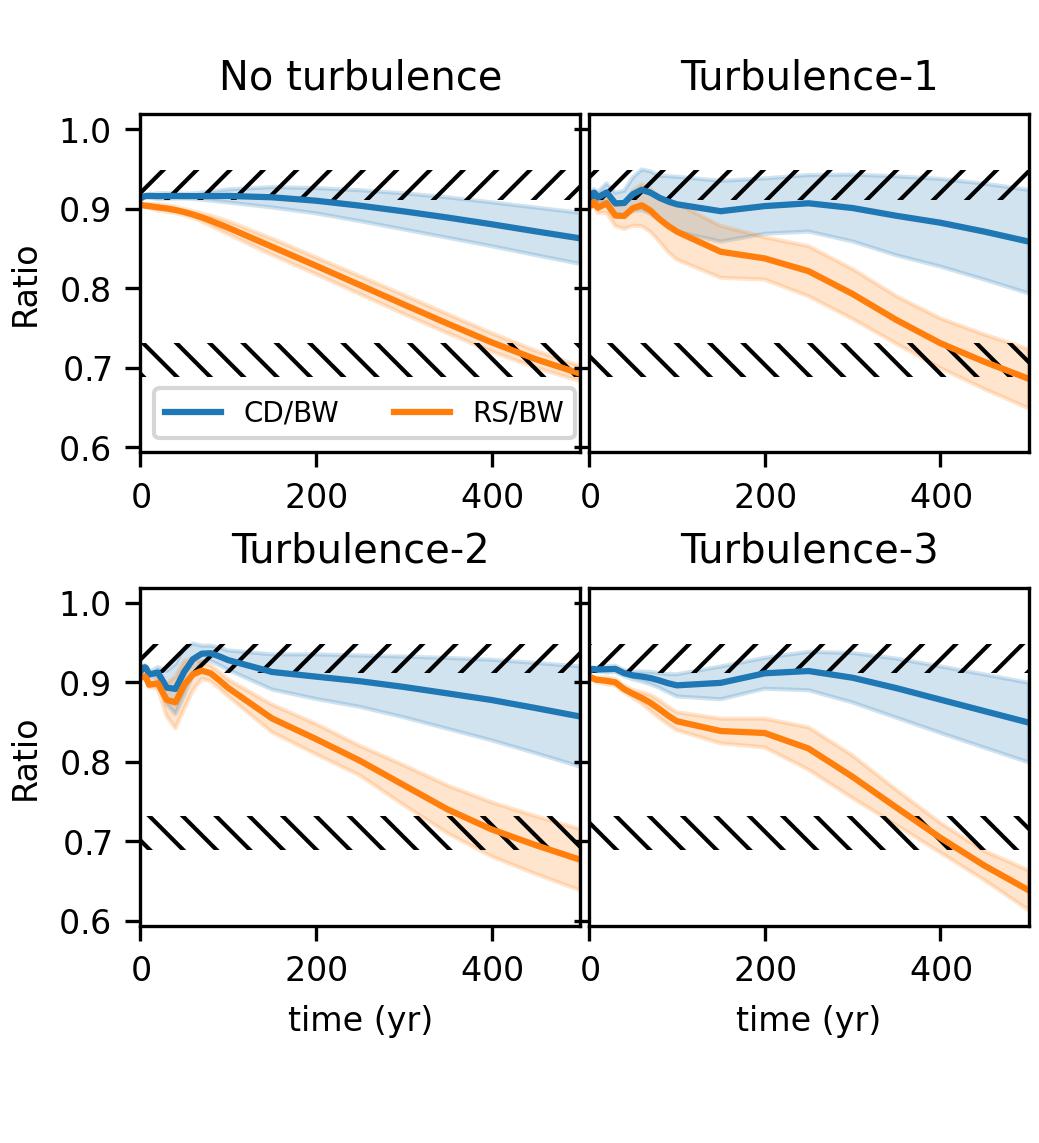} 
}
\subfigure[Evolution of the RTI width\label{fig:RTIb}]{
\includegraphics[]{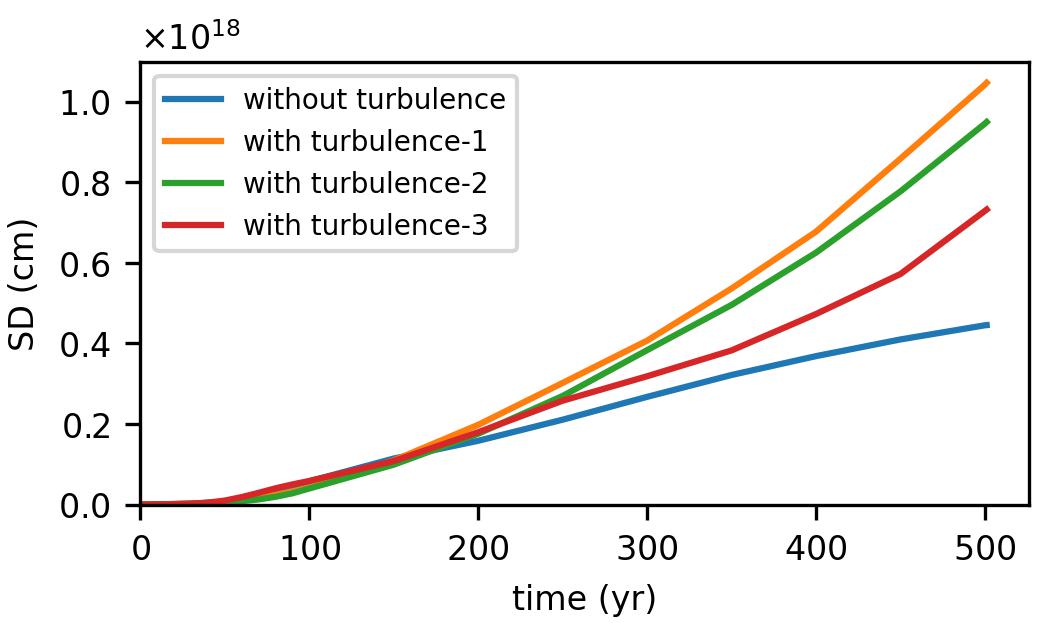}
}
\caption{Effect of the turbulence random phase set on the results of the simulations. The simulations are based on the piston-like profile (the same initial profile as the main article). As can be seen, the specific set of random phase influences the global evolution of the SNR without changing the global trend (RTI growth and radii ratios).}
\end{figure}

Here are some additional plots showing the effect of the turbulence on the BW:CD:RS radii ratios (figures \ref{fig:TEvolb}, \ref{fig:TEvolt}) as well as the RTI growth (figures \ref{fig:RTIb}, \ref{fig:RTIt}). The number that indexes the turbulence plot corresponds to a different set of random phases used to initialise the turbulence in the simulation.

\begin{figure}
\centering
\subfigure[Evolution of the ratios of radii\label{fig:TEvolp}]{
\includegraphics[]{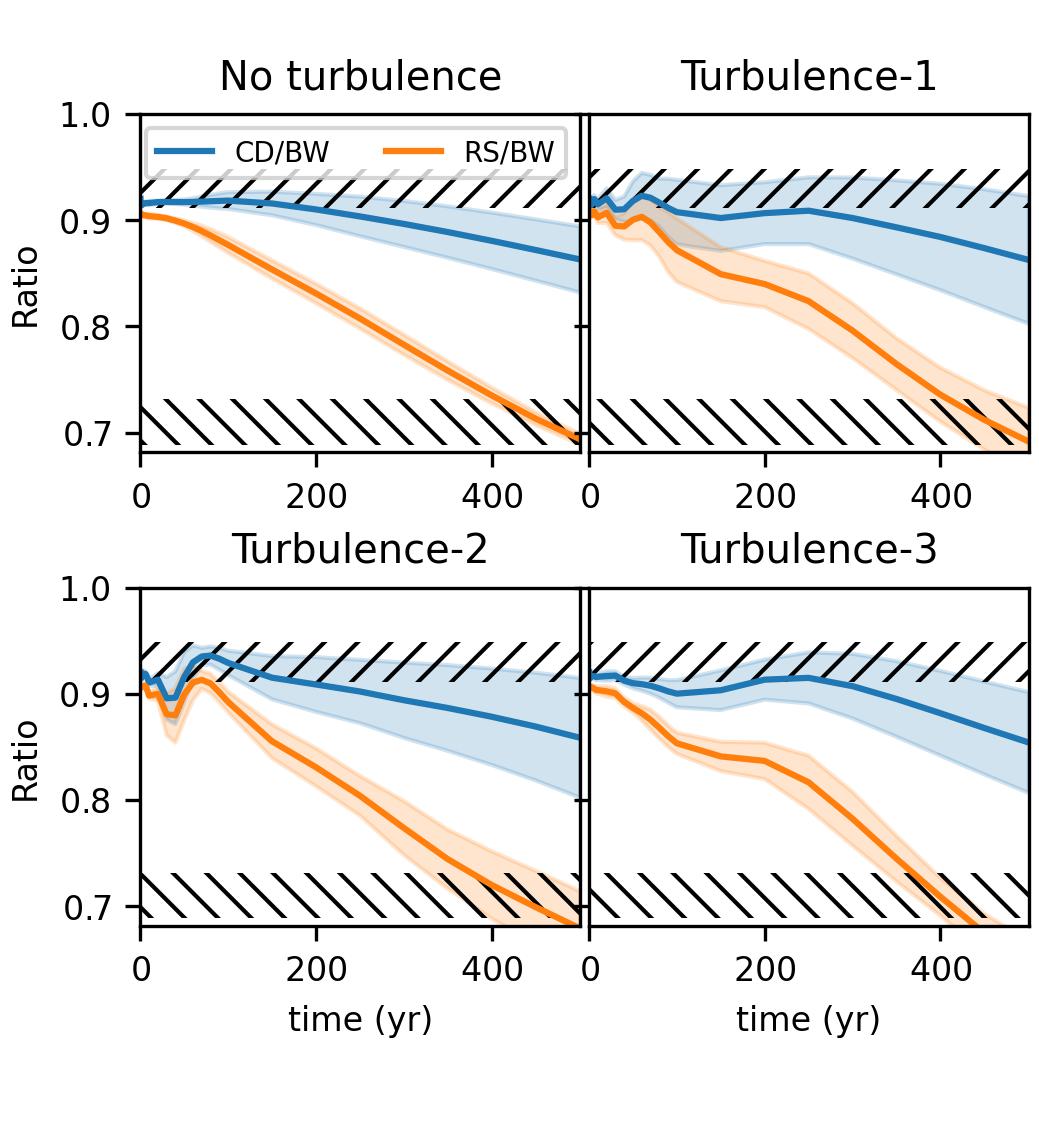} 
}
\subfigure[Evolution of the RTI width\label{fig:RTIp}]{
\includegraphics[]{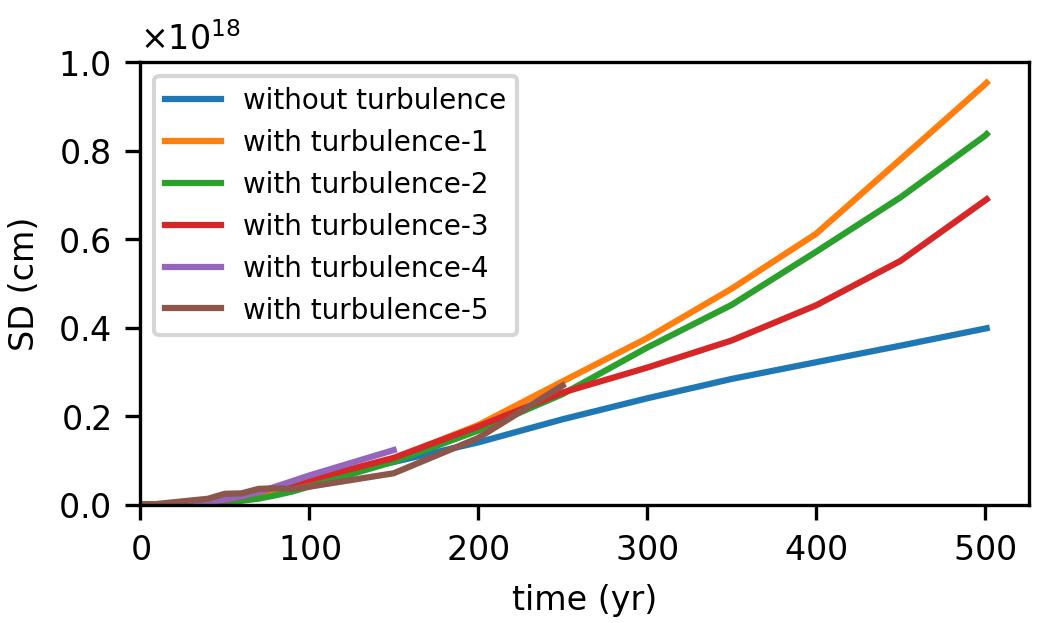}
}
\caption{Effect of the turbulence phase of a low-temperature ISM on the evolution of SNR. Here the simulation used the same parameters as the article: a piston-like profile for the SNR and same sets of phases for the turbulence. Only the temperature of the ISM differs, here \SI{10}{\K}. ISM temperature has a limited effect on the system.}
\end{figure}

\begin{figure}
\centering
\subfigure[Evolution of the ratios of radii\label{fig:TEvolc}]{
\includegraphics[]{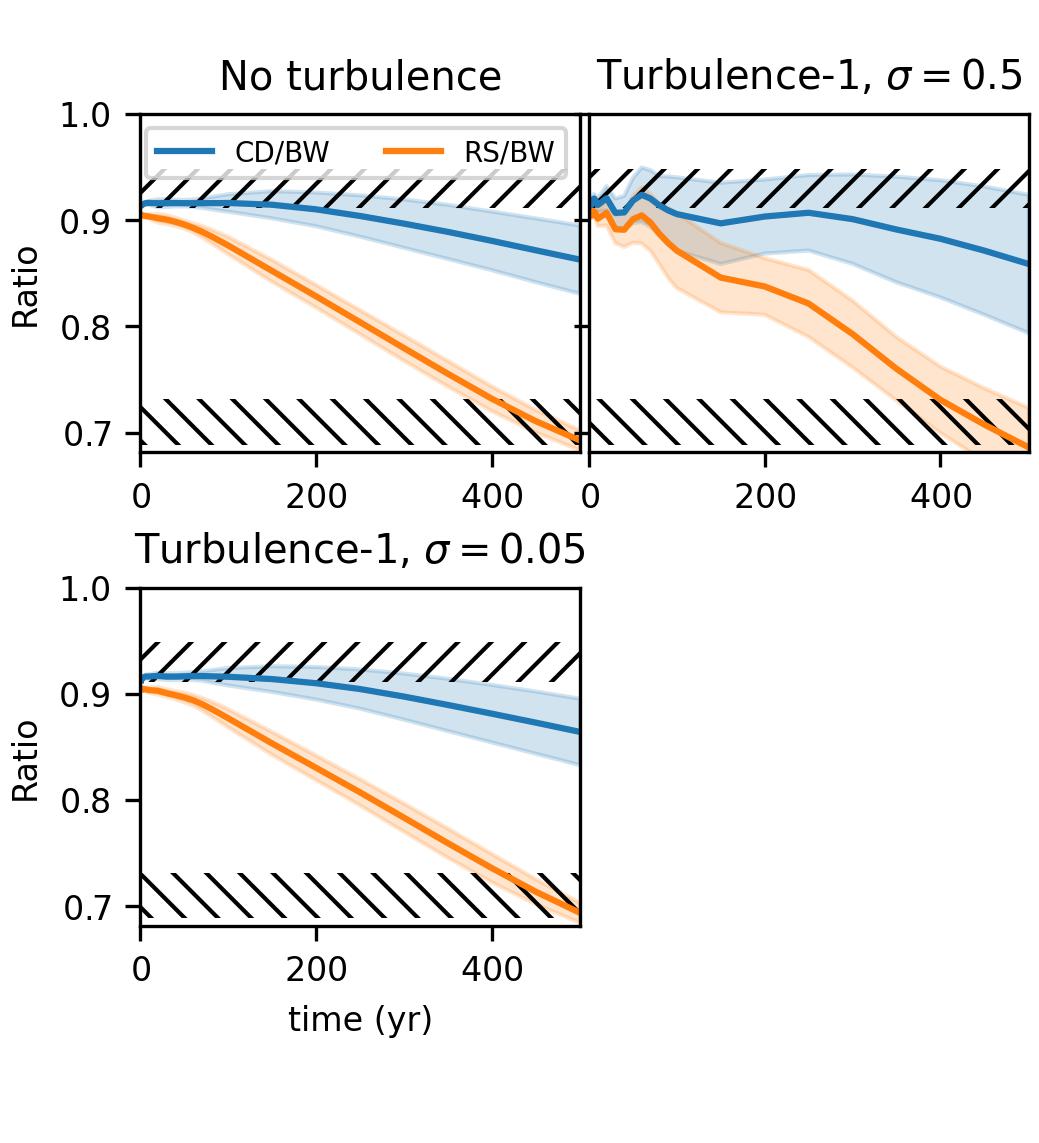} 
}
\subfigure[Evolution of the RTI width\label{fig:RTIc}]{
\includegraphics[]{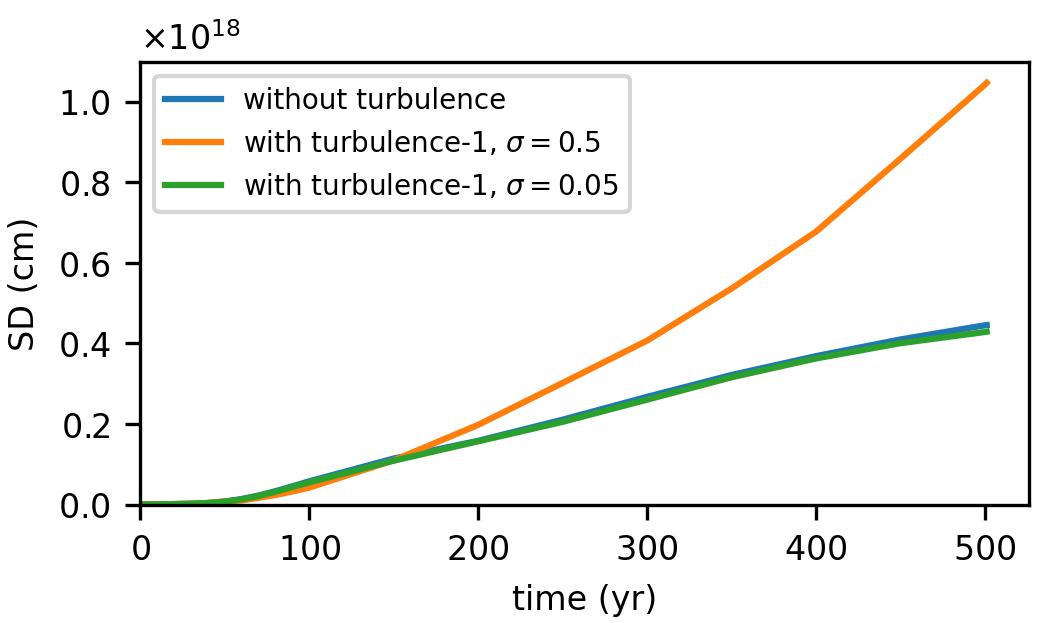}
}
\caption{Effect of the strength of the turbulence on the evolution of SNR. Here the simulation used the same parameters as in the article, except for the strength of the turbulence which includes a lower-$\sigma$ value (0.05) than the one used in the article (0.5). The strength of the turbulence is important for our conclusions.}
\end{figure}

\begin{figure}
\centering
\subfigure[Evolution of the ratios of radii. The dashed lines delimit the zone where both ratios are recreated at the same time. \label{fig:TEvolt}]{
\includegraphics[]{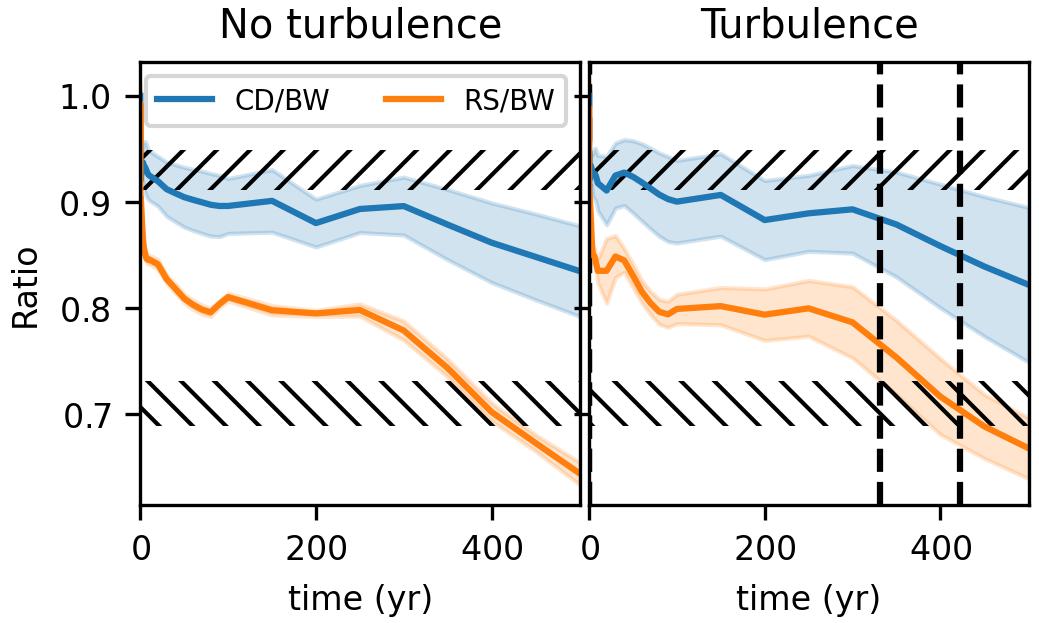} 
}
\subfigure[Evolution of the RTI width\label{fig:RTIt}]{
\includegraphics[]{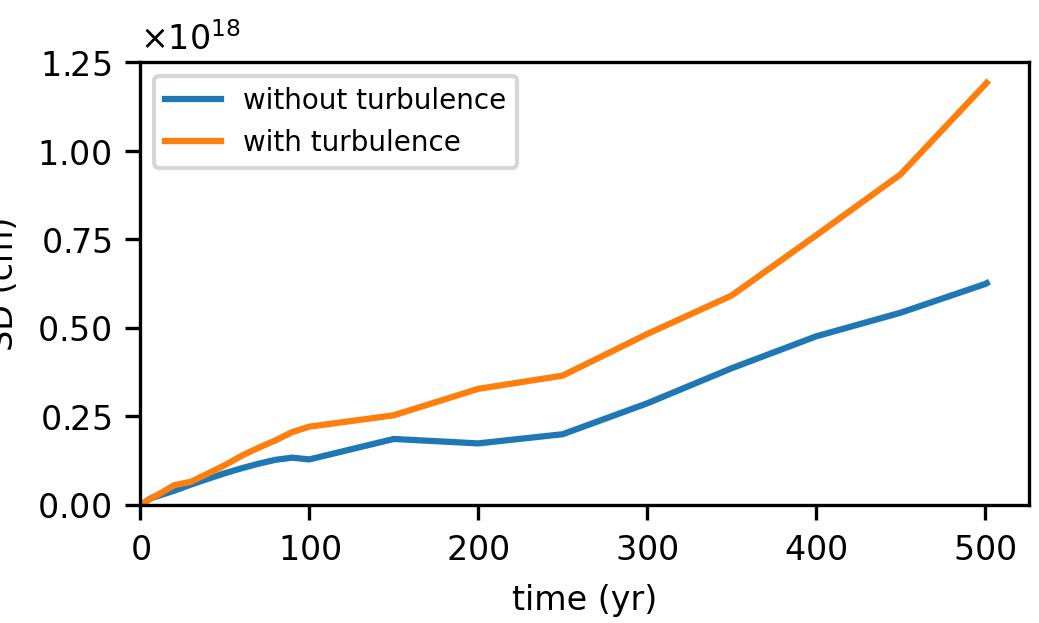}
}
\caption{Effect of the initial SNR profile on the simulation results. Contrary to the rest of the article, the simulations here are based on the DDTc profile taken from \cite{warren_cosmicray_2005}. While the specific dynamic differs from our piston-like profile, both the enhancement of RTI growth on the radii's ratios are recreated.}
\end{figure}
\end{appendix}
\end{document}